\documentclass[useAMS,usenatbib,usegraphicx]{mn2e}

\usepackage{amsmath,amssymb,amsbsy,amsfonts}
\usepackage{amssymb}
\usepackage[british]{babel}
\usepackage{psfrag}
\usepackage[squaren,thinspace,thinqspace]{SIunits}
\usepackage{mdwmath}

\newcommand{\clight}{\ensuremath{\mathsf{c}}}

\title{Formation of internal shock waves in bent jets}
\author[S. Mendoza \& M.S. Longair]
       {S. Mendoza\(^{1}\) \& M.S. Longair\(^2\)\\
               \(^1\) Instituto de Astronom\'{\i}a, Universidad Nacional 
	       Aut\'onoma de M\'exico, AP 70-264, Distrito Federal 04510,
	       M\'exico\\
               \(^2\) Cavendish Laboratory, Madingley Rd., Cambridge CB3 OHE,
	       U.K.}

\pagerange{\pageref{firstpage}--\pageref{lastpage}}

\begin{document}

\label{firstpage}

\maketitle

\begin{abstract}
  We discuss the circumstances under which the bending of a jet can
generate an internal shock wave.  The analysis is carried out for
relativistic and non--relativistic astrophysical jets.  The calculations
are done by the method of characteristics for the case of steady simple
waves.  This generalises the non--relativistic treatment first used
by \citet{icke91}.  We show that it is possible to obtain an upper
limit to the bending angle of a jet in order not to create a shock
wave at the end of the curvature.  This limiting angle has a value
of \( \sim \unit{75}{\degree}  \) for non--relativistic jets with
a polytropic index \( \kappa = 4/3 \), \( \sim \unit{135}{\degree}
\) for non--relativistic jets with \( \kappa = 5/3 \) and \( \sim
\unit{50}{\degree} \) for relativistic jets with \( \kappa = 5/3 \).
We also discuss under which circumstances jets will form internal shock
waves for smaller deflection angles.
\end{abstract}


\section{Introduction}

   In a previous article \citep{mendoza00}, we discussed a mechanism
by which galactic and extragalactic jets, can change their original
straight trajectory if they pass through a stratified cold and high
density region.  For the case of galactic jets this could be a cloud
in the vicinity of an H--H object.  For extragalactic jets, this region
could be a nearby galaxy, the interstellar medium of the host galaxy or
the intracluster medium itself.  Physically, what is important is that
the  medium interacting with the expanding jet from the source has a
non--uniform density.

  As mentioned by \citet{icke91} and  \citet{mendoza00}, when a jet
bends, it is in direct contact with the surroundings and one should
expect that entrainment from the surrounding gas might cause a severe
disruption of the jet itself.  Assuming that this entrainment is not
important, for example by an efficient cooling, what its left is a high
Mach number flow inside a collimated flow that bends.  When a supersonic
flow bends, the characteristics emanating at each point of the flow tend
to intersect \citep{daufm,courant}.  Since every hydrodynamical quantity
has a constant value on a given characteristic line, this intersection
causes the different physical quantities in the flow, such as the density
or pressure, to be multivalued.  This situation cannot occur in nature
and a shock wave is formed.

  The formation of shock waves inside a jet  are potentially dangerous.
These shock waves could give rise to subsonic flow in the jet and
collimation might no longer be achieved.  The present article discusses
the circumstances under which an internal shock  wave should be expected
in a bent jet.  The analysis presented in this article generalises
that made by \citet{icke91} by introducing relativistic effects into
the flow.  This generalisation has a drastic effect on the results.
Relativistic jets cannot bend as much as non--relativistic jets.  As we
will see below this occurs because, when relativistic effects are taken
into account, the characteristic lines in the flow are beamed in the
direction of the flow velocity.  The beaming increases as the velocity
of light is approached by the flow.  In other words, the chances for an
intersection between characteristic lines of uniform plane--parallel flow
(before the bending) and a curved flow increase because of this beaming.

  In order to analyse in detail the bending of relativistic jets we
discuss the following points in each section. We write down the equations
of relativistic hydrodynamics in section~\ref{basic-equations} and their
non--relativistic counterparts.  This sets the scene for discussing
the propagation of disturbances in flows which leads naturally to the
definition of the relativistic Mach number and characteristics.  This
generalises the traditional approach to gas dynamics and characteristic
surfaces such as that discussed for non--relativistic hydrodynamics
by \citet{daufm}.  The relativistic Mach number was first introduced
by \citet{chiu73} and some of its properties are well described by
\citet{konigl80}.  The most important result to be proved in this
section is the beaming of characteristic lines in the direction of
motion of the flow for relativistic flows. Section~\ref{prandtl-meyer}
analyses the relativistic and non--relativistic cases for a flow
depending on one angular variable, known as Prandtl--Meyer flow.
This leads naturally to the generalisation of a type of flow called
rarefaction waves in non--relativistic hydrodynamics \citep{daufm} which
in turn is essential for the understanding of flows that move through
a certain angle.  In section~\ref{steady-simple-waves} we use the main
results of the Prandtl--Meyer flow to describe steady simple waves.
These waves form naturally for steady plane parallel flow at infinity
when the jet turns through an angle with a certain curved profile.  It is
then possible to apply the necessary physical ingredients  to the steady
simple waves formed.  We then calculate an upper limit to the deflection
of a jet in order to avoid the formation of a shock at the end of its
curvature.  This limit does not mean that a jet which curves through a
small deflection angle is safe from generating an internal shock wave.
So, we also calculate a lower limit, for which a shock could form at
the onset of the curvature.  Finally, in section~\ref{discussion} we
discuss the astrophysical consequences of the results described above.

\section{Basic equations}
\label{basic-equations}

  In order to understand the formation of internal shock waves in a bent
jet it is necessary to understand some of the basic properties of the
gas dynamics of relativistic flows.

  The equations of motion for an ideal relativistic flow are described
by the 4--dimensional Euler's equation \citep{daufm}:

\begin{equation}
  \omega u^k \frac{ \partial u_i }{ \partial x^k } = \frac{ \partial p }{
    \partial x^i } - u_i u^k \frac{ \partial p }{ \partial x^k },
\label{eq.2.58}
\end{equation}

\noindent in which Latin indices take the values \( 0,\ 1,\ 2,\ 3 \). The
vector \( x^k = ( \clight t, \boldsymbol{r}) \), where \( \boldsymbol{r}
\) represents the three dimensional radius vector, \( \clight \) the speed
of light and \( t \) the time coordinate.  The Galilean metric \( g_{ik}
\) for flat space time is given by \( g_{00} \! = \! 1 \), \( g_{11}
\! = \! g_{22} \! = \! g_{33} \! = \! -1 \) and \( g_{ik} \! = \! 0 \)
when \( i \! \neq k \!  \).  The pressure is represented by \( p \),
\( \omega = e + p \) is the enthalpy per unit proper volume and \( e \)
is the internal energy density.  The four--velocity \( u^k = \mathrm{d}
x^k / \mathrm{d}s \) where the relativistic interval \( \mathrm{d}s \)
is given by \( \mathrm{d}s^2 = g_{kl} \mathrm{d}x^k \mathrm{d}x^l \).
The values of the different thermodynamic quantities are measured in
their local proper frame.

  The space components of eq.(\ref{eq.2.58}) give the relativistic
Euler equation:

\begin{equation}
  \frac{ \gamma \omega }{ \clight^2 } \left\{ \frac{ \partial
    \boldsymbol{v} }{ \partial t } + \boldsymbol{v}
    \cdot \boldsymbol{\mathrm{grad}} \, \boldsymbol{v} \right\}
    = -\boldsymbol{\mathrm{grad}} \, p - \frac{ \boldsymbol{v}
    }{ \clight^2 } \frac{ \partial p }{ \partial t} ,
\label{eq.2.58b}
\end{equation}

\noindent where \( \boldsymbol{v} \) is the three dimensional flow
velocity and \( \gamma \) is the Lorentz factor for a flow with this
velocity.

  For an ideal flow, in the absence of sources and sinks, the relativistic
continuity equation is given by \citep{daufm}:

\begin{equation}
  \frac{ \partial n^k }{ \partial x^k } = 0,
\label{eq.2.54}
\end{equation}

\noindent where the particle flux 4--vector \( n^k \! = \! n u^k \) and
the scalar \( n \) is the proper number density of particles in the fluid.

  A polytropic gas obeys the relation:

\begin{equation}
  p \propto n^{\kappa},
\label{eq.2.89}
\end{equation}

\noindent where the \emph{polytropic index} \( \kappa \) is a constant
and has the value \( 5/3 \) for an adiabatic monoatomic gas in which
relativistic effects are not taken into account. In the case of an
ultrarelativistic photon gas it has a value of \( 4/3 \).  It is not
difficult to show that for a polytropic gas, the speed of sound \(
a \) and the enthalpy per unit mass, the \emph{specific enthalpy},
\( w \) of are related to each other by the following formula
\citep{stanyuokovich60}:

\begin{equation}
  \frac{ \clight^2 }{ w } = 1 - \frac{ 1 }{ \kappa - 1 } \frac{ a^2
    }{ \clight^2 }.
                                                 \label{eq.2.92}
\end{equation}

  The quantities in the relativistic case are defined with respect
to the proper system of reference of the fluid, whereas in  classical
mechanics these quantities are referred to the laboratory frame.  In the
relativistic case the thermodynamic quantities, such as the internal
energy density \( e \), the entropy density \( \sigma \) and the enthalpy
density \( \omega \) are all defined with respect to the proper volume
of the fluid.  In non--relativistic fluid dynamics, these quantities
are defined in units of the mass of the fluid element they refer to.
For instance, the specific internal energy \( \epsilon \), the specific
entropy \( s \) and the specific enthalpy \( w \) are all measured per
unit mass in the laboratory frame.  When taking the limit in which the
speed of light \( c \) tends to infinity we must also bear in mind that
the internal energy density \( e \) includes the rest energy density \(
n m \clight^2 \), where \( m \) is the rest mass of the particular fluid
element under consideration.  Therefore the following non--relativistic
limits should be taken in passing from relativistic to non--relativistic
fluid dynamics:

\begin{align}
  mn & \xrightarrow[\clight \rightarrow \infty]{} \rho \sqrt{ 1 -
    v^2/\clight^2 } \approx \rho \left( 1 - v^2 / 2 c^2 \right),
                                                        \notag         \\
  e & \xrightarrow[\clight \rightarrow \infty]{} nm\clight^2 + \rho
    \epsilon \approx \rho \clight^2 - \frac{1}{2} \rho v^2 + \rho \epsilon,
                                                        \notag          \\
  \frac{ \omega }{ n } & \xrightarrow[\clight \rightarrow \infty]{}
    m\clight^2 + m \left( \epsilon + \frac{ p }{ \rho } \right) \approx m \left(
    \clight^2 + w \right),
\label{eq.2.60}  
\end{align}

\noindent where \( \rho \) is the mass density of the fluid in the
laboratory frame.

\section{Characteristics and Mach number}
\label{characteristics}

  The properties of \emph{subsonic} and \emph{supersonic} flow,
are completely different in nature.  To begin with, let us see how
perturbations with small amplitudes are propagated along the flow for
both subsonic and supersonic flows.  For simplicity in the following
discussion we consider two dimensional flow only.  The relations obtained
below are easily obtained for the general case of three dimensions.

\begin{figure}
  \begin{center}
    \psfrag{F}{ \( \theta \) }
    \psfrag{H}{ \( \boldsymbol{v} \) }
    \psfrag{K}{ \( a \hat{ \boldsymbol{e} }_r' \) }
    \psfrag{L}{ (a) }
    \psfrag{M}{ (b) }
    \psfrag{X}{ \( \boldsymbol{v} \) }
    \psfrag{Y}{ \( a \hat{ \boldsymbol{e} }_r' \) }
    \psfrag{Z}{ \( \boldsymbol{u} \) }
    \psfrag{A}{ \( \alpha \) }
    \psfrag{O}{ \( 0 \) }
    \includegraphics[scale=0.50]{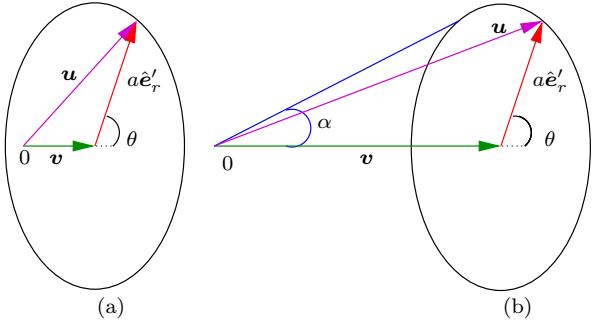}
  \end{center}
  \caption[Region of influence of small amplitude perturbations]{ Region
           of influence of small amplitude perturbations.  A perturbation
           of small amplitude is produced in the flow at some point  \(
           0 \).  This is carried by the flow which has velocity \(
           \boldsymbol{v} \).  In the case of subsonic flow, as shown
           in panel (a), the perturbation is able to propagate to the
           whole space.  When the flow is supersonic the perturbation is
           propagated only downstream inside a cone with aperture angle
           \( 2 \alpha \).  The speed of sound \( a \) and the angle \(
           \theta \) are measured in a frame of reference in which the
           flow is at rest --the proper frame of the flow.  The vector \(
           \boldsymbol{u} \) is the relativistic addition of vectors \(
           \boldsymbol{ v } \text{ and } a \hat{ \boldsymbol{ e } }_r'
           \), where \( \hat{ \boldsymbol{e} }_r' \) is a unit radial
           vector in the proper frame of the flow. }
\label{fig.2.1}
\end{figure}

  If a gas in steady motion receives a small perturbation, this propagates
through the gas with the velocity of sound relative to the flow itself.
In another system of reference, the \emph{laboratory frame}, in which the
velocity of the flow is \( v  \) along the \( x \) axis, the perturbation
travels with an observed velocity \( \boldsymbol{u} \) whose \( x \text{
and } y \) components are given by:

\begin{gather}
  u_x = \frac{ a \cos\theta + v }{ 1 + a v \cos\theta /
    \clight^2 },
                                                      \label{eq.2.81} \\
  u_y = \frac{ \gamma^{-1} a \sin\theta }{ 1 + a v \cos\theta /
    \clight^2 },
                                                      \label{eq.2.82}
\end{gather}

\noindent according to the rule for the addition of velocities in
special relativity \citep{daufields}.  The polar angle \( \theta \)
and the velocity of sound \( a \) are both measured in the proper frame
of the fluid.  Since a  small disturbance in the flow moves with the
velocity of sound in all directions, the parameter \( \theta \) can have
values  \( 0 \! \le \! \theta \! \le \! 2 \pi \).  This is illustrated
pictorially in Fig.~\ref{fig.2.1}.

  Let us consider first the case in which the flow is subsonic, as
illustrated in case (a) of Fig.~\ref{fig.2.1}.  Since by definition
\( v \! < \! a \) and \( \clight^2 > a v \), it follows from
eqs.(\ref{eq.2.81})-(\ref{eq.2.82}) that \( u_x( \theta \!  = \!
\pi ) \! < \! 0 \text{ and } u_y( \theta \! = \! \pi ) \! = \! 0 \).
In other words, the region influenced by the perturbation contains the
velocity vector \( \boldsymbol{v} \).  This means  that the perturbation
originating at \( 0 \) is able to be transmitted to all parts of the flow.

  When the velocity of the flow is supersonic, the situation is quite
different, as shown in case (b) of Fig.~\ref{fig.2.1}.  For this case, \(
u_x( \theta \! = \! \pi ) \! > \! 0 \text{ and } u_y( \theta \! = \! \pi
) \! = \! 0 \).  In other words, the velocity vector \( \boldsymbol{v}
\) is not fully contained inside the region of influence produced by
the perturbation.  This implies that only a bounded region of space
will be influenced by the perturbation originating at position \( 0 \).
For the case of steady flow, this region is evidently a cone.  Thus,
a disturbance arising at any point in supersonic flow is propagated only
downstream inside a cone of aperture angle \( 2 \alpha \).  By definition,
the angle \( \alpha \) is such that it is the angle subtended by the unit
radius vector \( \hat{ \boldsymbol{e} }_r \) with the velocity vector \(
\boldsymbol{ v } \) at the point in which the azimuthal unit vector \(
\hat{ \boldsymbol{e} }_\alpha \) is orthogonal with the tangent vector
\( \mathrm{d} ( a \hat{ \boldsymbol{e} }_r' ) / \mathrm{d} \theta \)
to the boundary of the region influenced by the perturbation.  The unit
vector \( \hat{ \boldsymbol{e} }_r' \) is the unit radial vector in
the proper frame of the flow.  In other words, the angle \( \alpha \)
obeys the following mathematical relation:

\begin{equation}
  \frac{ \mathrm{d} \left( a \hat{ \boldsymbol{e} }_r' \right) }{
  \mathrm{d} \theta } \cdot \hat{ \boldsymbol{e} }_\alpha = 0.
\label{eq.2.83}
\end{equation}

  Substitution of eqs.(\ref{eq.2.81})-(\ref{eq.2.82}) into
eq.(\ref{eq.2.83}) gives:

\begin{equation}
  \tan \alpha = - \gamma \left\{ \frac{ 1 }{ \tan \theta } + \frac{ a v
    }{ \clight^2 \sin \theta } \right\} .
\label{eq.2.84}
\end{equation}

\noindent On the other hand, since \( \tan \alpha \! = \! u_y / u_x \),
it follows from eqs.(\ref{eq.2.81})-(\ref{eq.2.82}) and eq.(\ref{eq.2.84})
that:

\[
  \tan \theta = - \frac{ v }{ a } \sqrt{ 1 - \frac{ a^2 }{ v^2 } }\, ,
\]

\noindent and so eq.(\ref{eq.2.84}) gives a relation between the angle \(
\alpha \), the velocity of the flow \( v \) and its sound speed \( a \):

\begin{equation}
  \tan \alpha = \gamma^{-1} \frac{ a / v }{ \sqrt{ 1 - \left( a / v
    \right)^2 } } 
\label{eq.2.85}
\end{equation}

\begin{figure}
  \begin{center}
    \psfrag{A}{ \( 2 \alpha \) }
    \psfrag{B}{ \( \boldsymbol{v} \) }
    \includegraphics[scale=0.78]{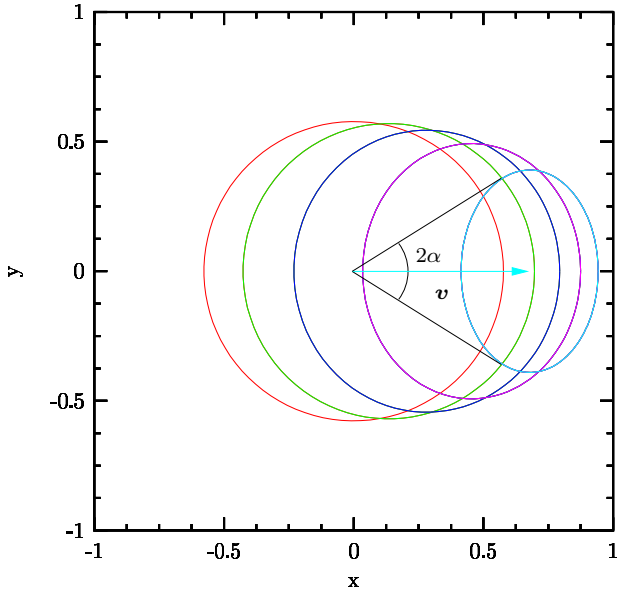}
  \end{center}
  \caption[Transmission of perturbations in supersonic flow]{ Region
           of influence of a perturbation for different values of the
           velocity of a relativistic gas with a sound speed \( a \! =
           \! \clight / \sqrt*{3} \).  From left to right the closed loops
           correspond to values of the velocity \( v \) of \( 0.0,\
           0.2,\ 0.4,\ldots ,0.8 \) in units of the speed of light \(
           \clight \).  The perturbation is assumed to originate at
           the origin of the proper system of reference of the flow.
           In the case of supersonic flow, the region of influence
           occurs only downstream inside a cone with aperture angle \(
           2 \alpha \).  This cone surrounds the corresponding loop and
           is tangent to it.  When the flow is subsonic, the perturbation
           is transmitted to all the flow. As a particular example, the
           cone and the velocity vector have been drawn for the case in
           which the velocity is \( v \! = \!  0.8 \clight\). }
\label{fig.2.2}
\end{figure}

  This variation of the angle \( \alpha \) with the velocity of the flow
is plotted in Fig.~\ref{fig.2.2} for the case in which the gas is assumed
to have a relativistic equation of state, that is, when \( p \! = \! e /
3 \).  The important feature to note from the plot is that the aperture
angle of the cone of influence is reduced when the velocity of the flow
approaches that of light.

  From eq.(\ref{eq.2.85}) it follows that, as the velocity of the
flow approaches that of light, the angle \( \alpha \) vanishes.
In other words, as the velocity reaches its maximum possible value, the
perturbation is communicated to a very narrow region along the velocity
of the flow.

  In studies of supersonic motion in fluid mechanics it is very useful
to introduce the dimensionless quantity \( M \) defined as:

\begin{equation}
  \frac{ 1 }{ M } \equiv \sin \alpha = \frac{ \gamma_a }{ \gamma }
    \frac{ a }{ v },
\label{eq.2.86}
\end{equation}

\noindent according to eq.(\ref{eq.2.85}). The quantity \( \gamma_a \! =
\! 1 / \sqrt{ 1 - ( a / \clight )^2 } \) is the Lorentz factor calculated
with the velocity of sound \( a \).   The number \( M \) has the property
that \( M \! \rightarrow \! 1 \text{ as } v \! \rightarrow \! a \text{ and
} M \! \rightarrow \! \infty \text{ as } v \! \rightarrow \! \clight \).
It also follows that \( M \! > \! 1 \) if and only if \( v \! > \! a \).

  The surface bounding the region reached by a disturbance starting
from the origin \( 0 \) is called a \emph{characteristic surface}
\citep{daufm}.  In the general case of arbitrary steady flow, the
characteristic surface is no longer a cone.  However, exactly as it
was shown above, the characteristic surface cuts the streamlines at any
point at the angle \( \alpha \).

  Let us briefly discuss the non--relativistic limit of the different
physical circumstances presented above.  To do this, we use the relations
in eq.(\ref{eq.2.60}) with \( \clight \! \rightarrow \! \infty \) and,
as usual for the non--relativistic case, we represent the speed of sound
by \( c \).

  The dimensionless number \( M \) satisfies the following relation:

\begin{equation}
   \frac{ 1 }{ M } = \sin{\alpha} \xrightarrow[\clight \rightarrow
     \infty]{} \frac{ c }{ v }
\label{eq.2.88}
\end{equation}

\noindent and is called in non--relativistic hydrodynamics the \emph{Mach
number}.\footnote{The relativistic generalisation of the Mach number as
presented in eq.\eqref{eq.2.86} was first calculated by \citet{chiu73},
who reduced the problem of steady relativistic gas dynamics to an
equivalent non--relativistic flow.  From eq.\eqref{eq.2.86} it follows
that this number is in fact a definition of the \emph{proper Mach
number} since it is defined as the ratio of the space component of
the relativistic four--velocity of the flow \( \gamma v \) to the same
component of the relativistic four--velocity of sound \( \gamma_a a \)
\citep{konigl80}.}

  The results obtained concerning the relativistic and non--relativistic
Mach number \( M \) can be rewritten in the following way: the
dimensionless Mach number \( M \) increases without limits as the
velocity of the flow takes its maximum possible value.  This maximum
value is the speed of light in the relativistic case and infinity in the
non--relativistic one.  The Mach number tends to zero as the velocity
of the flow vanishes, and tends to unity as the velocity of the flow
tends to the velocity of sound.  The Mach number is greater than one
for supersonic flow and less than unity when the velocity of the flow
is subsonic in both, the relativistic and non--relativistic cases.

\section{Prandtl--Meyer flow}
\label{prandtl-meyer}

  Let us describe briefly the exact solution of the equations of
hydrodynamics for plane steady flow which depends only on the angular
variable \( \phi \) only.  This problem was first investigated by Prandtl
and Meyer in 1908 \citep{daufm,courant} for the case in which relativistic
effects were not taken into account.  The full relativistic solution to
the problem is due to \citet{kolosnitsyn84}.

  For this case, Euler's equation, eq.(\ref{eq.2.58}) and the continuity 
equation, eq.(\ref{eq.2.54}) can be written:

\begin{gather}
  v_\phi = \frac{ \mathrm{d} v_r }{ \mathrm{d} \phi },
  							\label{eq.4.1} \\
    v_r n \gamma + \frac{ \mathrm{d} \left( v_\phi n \gamma \right) }{
    \mathrm{d} \phi } = 0,
    							\label{eq.4.2} \\
    \omega \gamma / n = \text{const},
    							\label{eq.4.3} 
\end{gather}

\noindent where \( v_r \text{ and } v_\phi \) are the components of
the velocity in the radial and azimuthal directions respectively.
Eq.(\ref{eq.4.3}) is the relativistic Bernoulli equation \citep{daufm}
for this problem.

  Using the definition of the speed of sound, 

\begin{equation}
  a = \clight \left( \frac{ \partial p }{  \partial e } \right)_\sigma 
\label{eq.2.69}
\end{equation}

\noindent  in eq.(\ref{eq.4.2}) and eq.(\ref{eq.4.3}) it is
found that:

\begin{equation}
  v_r + \frac{ \mathrm{d} v_\phi }{ \mathrm{d} \phi } + v_\phi \left(
   1 - \frac{ a^2 }{ \clight^2 } \right) \frac{ \clight^2 }{ a^2 } \frac{
   \mathrm{d} }{ \mathrm{d} \phi } \ln \left( \omega / n \right)  = 0.
\label{eq.4.4}
\end{equation}

\noindent On the other hand, differentiation of \( \gamma^{-2} \) with
respect to the azimuthal angle \( \phi \) and using eq.(\ref{eq.4.3}),
gives:

\begin{equation}
  v_\phi \left( v_r + \frac{ \mathrm{d} v_\phi }{ \mathrm{d} \phi }
    \right) + \clight^2 \left( 1 - \frac{ v_r^2 + v_\phi^2 }{ \clight^2
    } \right) \frac{ \mathrm{d} }{ \mathrm{d} \phi } \ln \left( \omega /
    n \right) = 0.
\label{eq.4.5}
\end{equation}

\noindent Multiplication of eq.(\ref{eq.4.4}) by \( v_\phi \) and
substracting this from eq.(\ref{eq.4.5}) gives:

\begin{equation}
  v_\phi^2 = a^2 \left( 1 - \frac{ v_r^2 }{ \clight^2 } \right).
\label{eq.4.6}
\end{equation}

  Bernoulli's equation, eq.(\ref{eq.4.3}), together with the value of
the specific enthalpy for a polytropic gas given in eq.(\ref{eq.2.92}),
can be rewritten

\begin{equation}
  \left( 1 - \frac{ v_r^2 + v_\phi^2 }{ \clight^2 } \right) \left( 1 -
    \frac{ 1 }{ \kappa - 1 } \frac{ a^2 }{ \clight^2 } \right) ^2 = 
    \left( 1 - \frac{ 1 }{ \kappa -1 } \frac{ a_0^2 }{ \clight^2 }
    \right)^2,
\label{eq.4.7}
\end{equation}

\noindent in which it has been assumed that at some definite point,
the flow velocity vanishes and the speed of sound has a value \( a_0
\) there.  It is always possible to make the velocity zero at a certain
point by a suitable choice of the system of reference.

  Eqs.(\ref{eq.4.6})-(\ref{eq.4.7}) can be solved in terms of \( v_r
\text{ and } v_\phi \):

\begin{gather}
  v_r^2 / \clight^2 = 1 - F^2(a),
  						\label{eq.4.8} \\
  v_\phi^2 = a^2 F^2(a),
  						\label{eq.4.9}  \\
  \intertext{where}
  F^2(a) = \left( 1 - \frac{ 1 }{ \kappa - 1 } \frac{ a_0^2 }{ \clight^2
    } \right)^2 \left( 1 - \frac{ a^2 }{ \clight^2 } \right)^{-1} \left(
    1 - \frac{ 1 }{ \kappa - 1 } \frac{ a^2 }{ \clight^2 } \right)^{-2}.
    						\label{eq.4.10} 
\end{gather}

  Because \( v_r \mathrm{d} v_r = \clight^2 F(a) F'(a) \mathrm{d}
a \), eq.(\ref{eq.4.1}) gives the required solution \citep{kolosnitsyn84}:

\begin{equation}
  \phi + \phi_0 = \pm \clight \int{ \frac{ F'(a) \mathrm{d} a }{ a \sqrt{
    1 - F^2(a) } } }.
\label{eq.4.11}
\end{equation}

\noindent This equation gives the speed of sound as a function of the
azimuthal angle.  From eqs.(\ref{eq.4.8})-(\ref{eq.4.9}) it follows that
the radial and azimuthal velocities can be obtained as a function of the
same angle \( \phi \).  As a result, all the remaining hydrodynamical
variables can be found.  The sign in eq.(\ref{eq.4.11}) can be chosen
to be negative by measuring the angle \( \phi \) in the appropriate
direction and we will do that in what follows.

  Let us consider now the case of an ultrarelativistic gas and integrate
eq.(\ref{eq.4.11}) by parts, to obtain:

\begin{equation}
  \phi + \phi_0 = \frac{ \clight }{ a } \arccos F(a) + \clight \int{
    \frac{ \mathrm{d} a }{ a^2 } \arccos F(a) }.
\label{eq.4.12}
\end{equation}

\noindent  For the case of an ultrarelativistic gas, the speed of sound
\( a \) is given by \citep{stanyuokovich60} \( a = \sqrt{ \kappa - 1 }
\ \clight \) .  In other words, this velocity is constant and so the
integral in eq.(\ref{eq.4.12}) is a Lebesgue integral.  Since this
integral is taken over a bounded and measurable function over a set of
measure zero, its value is zero.

  Using eqs.(\ref{eq.4.8})-(\ref{eq.4.9}) and eq.(\ref{eq.4.12}) the
desired solution is obtained \citep{kolosnitsyn84,konigl80}:

\begin{gather}
  v_r = \clight \sin \left\{ \sqrt{ \kappa - 1 } \left( \phi 
    + \phi_0 \right) \right\},
  						\label{eq.4.13} \\
  v_\phi = \sqrt{ \kappa - 1 } \ \clight \cos \left\{ \sqrt{ \kappa -
    1 } \left( \phi + \phi_0 \right) \right\},
						\label{eq.4.14}
\end{gather}

\noindent for an ultrarelativistic equation of state of the gas.

  For the non--relativistic case, in which \( \clight \! \rightarrow
\! \infty \), eq.(\ref{eq.4.11}) gives for a polytropic gas with
polytropic index \( \kappa \):

\begin{gather}
  \phi + \phi_0 = - \sqrt{ \frac{ \kappa + 1 }{ \kappa - 1 } }\int{
    \frac{ \mathrm{d} \zeta }{ \sqrt{ 1 - \zeta^2 } } },
						\notag \\
  \intertext{where}
  \zeta \equiv \frac{ a }{ \clight } \sqrt{ \frac{ \kappa + 1 }{ \kappa -
    1 } } \left\{ 1 - \left( 1 - \frac{ 1 }{ \kappa -1 } \frac{ a_0^2 }{
    \clight^2 } \right)^2 \right\}^{-1/2},
    						\notag \\
  \intertext{and so, the required solution is \citep{kolosnitsyn84}: } 
  \phi + \phi_0 = \sqrt{ \frac{ \kappa + 1 }{ \kappa - 1 } } \arccos
    \left( \frac{ c }{ c_* } \right),
						\label{eq.4.15}
\end{gather}

\noindent where the speed of sound \( a \) has been rewritten as \(
c \) to be consistent in the non--relativistic case. The \emph{critical
velocity of sound} \( c_* \) is given by \citep{daufm}:

\begin{equation}
  c_*^2 = \frac{ 2 }{ \kappa + 1 } c_0^2.
\label{eq.4.16}
\end{equation}

\noindent The value for the velocities can thus be calculated from
eq.(\ref{eq.4.1}) and eq.(\ref{eq.4.9})  with \( F(a) \! = \! 1 \):

\begin{gather}
  v_r = \sqrt{ \frac{ \kappa + 1 }{ \kappa - 1 } } c_* \sin \sqrt{
    \frac{ \kappa - 1 }{ \kappa +1 } } \left( \phi - \phi_0 \right),
						\label{eq.4.17} \\
  v_\phi = c = c_* \cos \sqrt{ \frac{ \kappa - 1 }{ \kappa +1 } } \left(
    \phi - \phi_0 \right).
						\label{eq.4.18}
\end{gather}

  Some important inequalities must be satisfied for the flow
under consideration.  First of all, eq.(\ref{eq.4.11}) together
with eq.(\ref{eq.2.92}) and the first law of thermodynamics, \(
\mathrm{d}\left( \omega / n \right) = T \mathrm{d} \left( \sigma /
n \right) + \left( 1 / n \right) \mathrm{d} p \) \citep{daufm},
imply that \( \mathrm{d} p / \mathrm{d} \phi \! < \! 0 \).  Using
this inequality and the fact that \( \mathrm{d} e \! = \! \clight^2
\mathrm{d} p / a^2 \) combined with the first law of thermodynamics,
it follows that \( \mathrm{d} n / \mathrm{d} \phi \! < \! 0 \).  Also,
using eqs.(\ref{eq.4.8})-(\ref{eq.4.9}) it follows that \( \mathrm{d}
v / \mathrm{d} \phi \propto - \mathrm{d} a / \mathrm{d} \phi \) and
necessarily \( \mathrm{d} v / \mathrm{d} \phi \! > \! 0 \).

\begin{figure}
  \begin{center}
    \psfrag{A}{ \( v_\phi \hat{ \boldsymbol{e} }_\phi  \) }
    \psfrag{B}{ \( \boldsymbol{v} \) }
    \psfrag{C}{ \( v_r \hat{ \boldsymbol{e} }_r \) }
    \psfrag{D}{ \( \phi \) }
    \psfrag{E}{ \( \chi \) }
    \psfrag{F}{ \( 0 \) }
    \psfrag{G}{ \( \alpha \) }
    \includegraphics[scale=0.83]{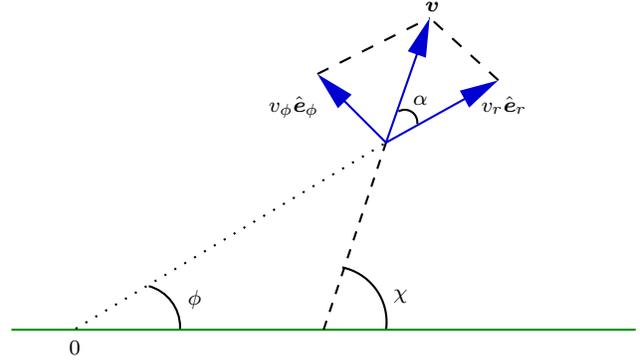}
  \end{center}
  \caption[Relation between the velocity vector and certain given
           direction]{ Relation between the velocity vector \(
           \boldsymbol{ v } \! = \!  v_r \hat{ \boldsymbol{e} }_r +
           v_\phi \hat{ \boldsymbol{e} }_\phi \) and the angle \(
           \chi \), as a function of the azimuthal angle \( \phi \).
           \( \chi \) is the angle that the velocity vector makes with
           certain fixed axis with origin O.}
\label{fig.1}
\end{figure}

  On the other hand, the angle \( \chi \) that the velocity vector makes
with some fixed axis is related to the velocity and the azimuthal angle \(
\phi \) by:

\begin{equation}
  \chi = \phi + \arctan \left( v_\phi / v_r \right)
\label{eq.4.19}
\end{equation}

\noindent as it shown in Fig.~\ref{fig.1}. Thus, since the \(
\phi \) component of Euler's equation, eq.(\ref{eq.2.58}) implies that:

\[
  \left( v_r + \frac{ \partial v_\phi }{ \partial \phi } \right) \frac{
    \gamma^2 v_\phi \omega }{ n } + \clight^2 \frac{ \partial \left(
    \omega / n \right) }{ \partial n} = 0,
\]

\noindent it follows that \( \mathrm{d} \chi / \mathrm{d} \phi \! =
\! - \left(  v^2 \gamma^2 \omega / \clight^2 \right)^{-1} \mathrm{d}
p / \mathrm{d} \phi \).

  In other words, we have proved that for the flow with which we are
concerned, the following inequalities are satisfied:

\begin{equation}
  \mathrm{d} p / \mathrm{d} \phi < 0, \quad \mathrm{d} n / \mathrm{d}
    \phi < 0 , \quad \mathrm{d} v / \mathrm{d} \phi > 0, \quad \mathrm{d}
    \chi / \mathrm{d} \phi > 0.
\label{eq.4.20}
\end{equation}

\noindent A flow  with these properties is described as a
\emph{rarefaction wave} in non--relativistic fluid dynamics \citep{daufm}
and we will use this name in what follows.

  Another, very important property of this rarefaction wave is
that the lines at constant \( \phi \) intersect the streamlines at
the Mach angle, that is, they are characteristics.  Indeed, from
Fig.~\ref{fig.1}, it follows that the angle \( \alpha \) between the
line \( \phi \! = \!  \textrm{const} \) and the velocity vector \(
\boldsymbol{ v } \) is given by \( \sin \alpha = v_\phi / v \).  Using
eqs.(\ref{eq.4.8})-(\ref{eq.4.10}) it follows that this relation can be
written as eq.(\ref{eq.2.86}).   Because all quantities in the problem
are functions of a single variable, the angle \( \phi \), it follows
that every hydrodynamical quantity is constant along the characteristics.

\section{Steady simple waves}
\label{steady-simple-waves}

  Let us now consider the two dimensional problem of steady plane
parallel flow which then turns through an angle as it flows round a
curved profile.  A particular case of this problem, when the flow turns
through an angle is described by \citet{daufm}.  For this particular
situation the Prandtl-Meyer flow is obviously the solution and so,
the hydrodynamical quantities depend on a single variable, the angle
\(\phi\) measured from a defined axis at the onset of the curvature.
Because of this, all quantities can be expressed as functions of each
other.  Since this case is a particular solution to the general problem,
it is natural to seek the solutions of the equations of motion in which
the quantities \( p,\ n,\ v_x,\ v_y \) can be expressed as a function of
each other.  Evidently this imposes a restriction on the solution of the
equations of motion since for two dimensional flow, any quantity depends
on two coordinates, \( x \) and \( y \), and so any chosen hydrodynamical
variable can be written as a function of any other two.

  Because of the fact that the flow is uniform at infinity, where all
quantities are constant, particularly the entropy, and because the flow
is steady, the entropy is constant along a streamline.  Thus, if there are
no shock waves in the flow, the entropy remains constant along the whole
trajectory of the flow and in what follows we will use this result.

  In this case, Euler's equation, eq.(\ref{eq.2.58b}), and the continuity
equation, eq.(\ref{eq.2.54}), are respectively:

\begin{gather*}
  v_x \frac{ \partial v_x }{ \partial x } + v_y \frac{ \partial v_x }{
    \partial y } = - \frac{ \clight^2 }{ \gamma \omega } \frac{ \partial
    p }{ \partial x},
							\\
  v_x \frac{ \partial v_y }{ \partial x } + v_y \frac{ \partial v_y }{
    \partial y } = - \frac{ \clight^2 }{ \gamma \omega } \frac{ \partial
    p }{ \partial y},
    							\\
  \frac{ \partial }{ \partial x } \left( \gamma v_x n \right) + \frac{
    \partial }{ \partial y } \left( \gamma v_y n \right) = 0.
\end{gather*}

\noindent Rewriting these equations as Jacobians\footnote{The Jacobian
\( \partial ( a, b ) / \partial ( x, y ) \) is defined as:
\[
  \frac{ \partial (a, b) }{ \partial (x,y) } = \textrm{det} 
       \begin{bmatrix}
         \partial a / \partial x & \partial a / \partial y \\
         \partial b / \partial x & \partial b / \partial y 
        \end{bmatrix}.
\]
} we obtain:

\begin{gather*}
  v_x \frac{ \partial ( v_x, y ) }{ \partial ( x, y ) } - v_y \frac{
    \partial ( v_x, x ) }{ \partial ( x, y ) } = - \frac{ \clight^2 }{
    \gamma \omega } \frac{ \partial ( p, y ) }{ \partial ( x, y ) },
    							\\
  v_x \frac{ \partial ( v_y, y ) }{ \partial ( x, y ) } - v_y \frac{
    \partial ( v_y, x ) }{ \partial ( x, y ) } = + \frac{ \clight^2 }{
    \gamma \omega } \frac{ \partial ( p, x ) }{ \partial ( x, y ) },
    							\\
  \frac{ \partial ( \gamma v_x n, y ) }{ \partial ( x, y ) } 
    - \frac{ \partial ( \gamma v_y n, x ) }{ \partial ( x, y ) } = 0.
\end{gather*}

\noindent We now take the coordinate \( x \) and the pressure \( p \) as
independent variables.  To make this transformation we have to multiply
the previous set of equations by \( \partial ( x, y ) / \partial ( x,
p ) \).  This multiplication leaves the equations the same, but with the
substitution \( \partial ( x, y ) \! \rightarrow \partial ( x, p ) \).
Expanding this last relation and because all quantities are now functions
of the pressure \( p \) but not of \( x \), it follows that:

\begin{gather*}
  \left( v_y - v_x \frac{ \partial y }{ \partial x } \right) \frac{
    \mathrm{d} v_x }{ \mathrm{d} p } = \frac{ \clight^2 }{ \gamma \omega
    } \frac{ \partial y }{ \partial x },
							\\
  \left( v_y - v_x \frac{ \partial y }{ \partial x } \right) \frac{
    \mathrm{d} v_y }{ \mathrm{d} p } = - \frac{ \clight^2 }{ \gamma \omega
    } \frac{ \partial y }{ \partial x },
							\\
   \left( v_y - v_x \frac{ \partial y }{ \partial x } \right) \frac{
    \mathrm{d} \left( \gamma n \right) }{ \mathrm{d} p } + \gamma n
    \left\{ \frac{ \mathrm{d} v_y }{ \mathrm{d} p } - \frac{ \partial
    y }{ \partial x } \frac{ \mathrm{d} v_x }{ \mathrm{d} p } \right\} =
    0.
\end{gather*}

  Here we have taken \( \partial y / \partial x \) to mean the derivative
at constant pressure: \( ( \partial y / \partial x )_p \).  Since every
hydrodynamic quantity is assumed to be a function of the pressure, then in
the previous set of equations it necessarily follows that \( \partial y /
\partial x \) is a function which depends only on the pressure, that is \(
( \partial y / \partial x )_p = f_1(p) \). Therefore:

\begin{equation}
  y = x f_1(p) + f_2(p).
\label{eq.4.21}
\end{equation}

  No further calculations are needed if we use the solution for the
case in which a rarefaction wave is formed when flow turns through
an angle \citep{daufm}.  This solution is given by the results of
section \ref{prandtl-meyer}.  As was mentioned in that section, all
hydrodynamical quantities are constant along the characteristic lines
\( \phi \! = \! \textrm{const} \).  The particular solution of the flow
past an angle obviously corresponds to the case in which \( f_2(p) \! =
\! 0 \) in eq.(\ref{eq.4.21}).  The function \( f_1(p) \) is determined
from the equations obtained in section \ref{prandtl-meyer}.

  For a given constant value of the pressure \( p \), eq.(\ref{eq.4.21}),
gives a set of straight lines in the \( x \text{--} y \) plane.  These
lines intersect the streamlines at the Mach angle.  This occurs because
the lines \( y \! = \!  x f_1(p) \) for the particular solution of the
flow through an angle have this property.  In other words, one family
of characteristic surfaces correspond to a set of straight lines along
which all quantities remain constant.  However, for the general case,
these lines are no longer concurrent.

  The properties of the flow described above are analogous to
the non--relativistic equivalent known as \emph{simple waves}
\citep{daufm}. In what follows we will use this name to refer to such
a flow.

\begin{figure}
  \begin{center}
    \psfrag{B}{ \( \alpha_1 \) }
    \psfrag{C}{ \( \phi_1 \) }
    \psfrag{D}{ \( \phi_* \) }
    \psfrag{E}{ \( \phi_* - \phi \) }
    \psfrag{Z}{ \( \theta \) }
    \includegraphics[scale=0.5]{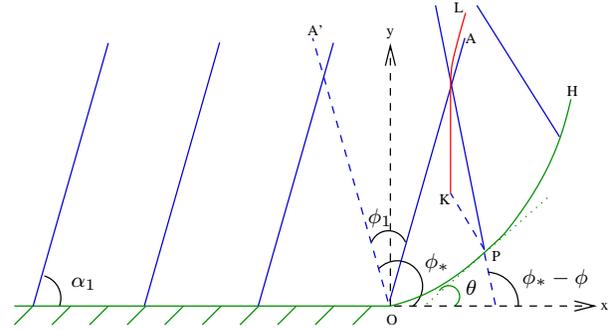}
  \end{center}
  \caption[Formation of shock waves in flows that turn through an
           angle]{Supersonic uniform flow at the left of the diagram bends
           around a curved profile OH.  The Mach angle \( \alpha_1 \)
           is the angle made by the characteristics and the streamlines
           before the onset of the curvature.  The characteristics
           make an angle \( \phi_1 \) with the ``continuation'' of
           the rarefaction wave formed at the onset of the curvature
           and the angle \( \phi \) is measured from the line 0A'.
           The curvature causes the characteristic lines to intersect
           eventually and this occurs at point K in the diagram, giving
           rise to a shock wave represented as the segment KL.}
\label{fig.2}
\end{figure}

  Let us now construct the solution for a simple wave once the curved
profile is fixed. Consider the profile as shown in Fig.~\ref{fig.2}.
Plane parallel steady flow streams in from the left of point O and flows
around the curved profile. Since we assume that the flow is supersonic,
the effect of the curvature starting at O is communicated to the flow
only downstream of the characteristic OA generated at point O. The
characteristics to the left of OA, \mbox{region 1}, are all parallel
and intersect the \( x \) axis at the Mach angle \( \alpha_1 \) given
by eq.(\ref{eq.2.86}):

\begin{equation}
  \sin \alpha_1 = \frac{ \sqrt{ 1 - \left( v_1 / \clight \right)^2 }
    }{ \sqrt{ 1 - \left( a / \clight \right)^2 } } \ \frac{ a }{ v_1 },
\label{eq.4.22}
\end{equation}

\noindent where the velocity \( v_1 \) is the velocity of the flow to the
left of the characteristic OA.  In eqs.(\ref{eq.4.11})-(\ref{eq.4.18})
the angle \( \phi \) of the characteristics is measured with respect
to some straight line in the \( x \text{--} y \) plane.  As a result,
we can choose for those equations the constant of integration \( \phi_0
\! \equiv 0 \!  \)\hskip2pt .  This means that the line from which the
angle \( \phi \) is measured has been chosen in a rather special way.
In order to find the line which is the characteristic for \( \phi \! = \!
0 \), let us proceed as follows.  When \( \phi \! = \! 0 \) and the gas
is ultrarelativistic, eqs.(\ref{eq.4.13})-(\ref{eq.4.14}) show that
the velocity \( v \! = \! a \) and for the non--relativistic case,
it follows from eqs.(\ref{eq.4.17})-(\ref{eq.4.18}) that the velocity
takes the value \( v \! = \!  c \).  In both cases this means that the
line \( \phi \! = \! 0 \) corresponds to the point at which the flow has
reached the value of the local velocity of sound.  This, however, is not
possible since we are assuming that the flow is supersonic everywhere.
Nevertheless, if the rarefaction wave is assumed to extend formally into
the region to the left of OA, we can use these relations and then the
characteristic line must correspond to a value of \( \phi \) given by:

\begin{gather}
  \phi_1 = \sqrt{ \frac{ \kappa + 1 }{ \kappa - 1 } } \arccos
    \left( \frac{ c_1 }{ c_* } \right),
    							\label{eq.4.23} \\  
  \intertext{for a non--relativistic gas according to eq.\eqref{eq.4.15}, and}
  \phi_1 = \frac{ \clight }{ a } \arccos \frac{ \sqrt{ 1 - ( v_1 /
    \clight )^2 } }{ \sqrt{ 1 - ( a / \clight )^2 } },
							\label{eq.4.24}
\end{gather}

\noindent for the ultrarelativistic case according to eq.\eqref{eq.4.7},
eq.\eqref{eq.4.10} and eq.(\ref{eq.4.12}).  The angle between the
characteristics and the \( x \) axis is then given by: \( \phi_* - \phi
\), where \( \phi_* \!  = \alpha_1 + \phi_1, \text{ and the angle }
\alpha_1 \) is the Mach angle in region 1.  The \( x \text{ and } y \)
velocity components in terms of the azimuthal angle \( \theta \) are
given by:

\begin{equation}
  v_x = v \cos \theta, \qquad  v_y = v \sin \theta,
\label{eq.4.25}
\end{equation}

\noindent and the values for the magnitude of the velocity, the angle \(
\theta \) and the pressure are given by:

\begin{gather}
  v^2 = c_*^2 \left\{ 1 + \frac{ 2 }{ \kappa - 1 } \sin^2 \sqrt{ \frac{
   \kappa - 1 }{ \kappa + 1 } } \phi \right\},
   							\label{eq.4.26} \\
  \begin{split}
    \theta 
      & = \phi_* - \phi - \alpha, \\
      & = \phi_* - \phi - \arctan \left\{ \sqrt{ \frac{ \kappa -1 }{
        \kappa + 1 } } \cot \sqrt{ \frac{ \kappa - 1 }{ \kappa + 1 } }
        \phi  \right\}, \\
   \end{split} 
   							\label{eq.4.27} \\
  p = p_* \cos^{ 2 \kappa / \left( \kappa - 1 \right) } \sqrt{ \frac{
    \kappa - 1 }{ \kappa + 1 } } \phi ,
							\label{eq.4.28}
\end{gather}

\noindent for a non--relativistic gas according to
eqs.(\ref{eq.4.15})-(\ref{eq.4.18}) and using the fact that the Poisson
adiabatic for a polytropic gas means that: \( p c^{- 2 \kappa / \left(
\kappa - 1 \right) } \! = \! \textrm{const} \). In the case of an
ultrarelativistic gas, eqs.(\ref{eq.4.12})-(\ref{eq.4.14}) together
with Bernoulli's equation and the fact that the enthalpy density \(
\omega \! = \! \kappa p / \left( \kappa - 1 \right) \) give:

\begin{gather}
  v^2 = \clight^2 \left\{ 1 - \left( 2 - \kappa \right) \cos^2 \sqrt{
    \kappa - 1 } \phi \right\},
  							\label{eq.4.29} \\
  \begin{split}
    \theta 
      & = \phi_* - \phi - \alpha, \\
      & = \phi_* - \phi - \arctan \left\{ \sqrt{ \kappa -1 } \cot \sqrt{
        \kappa - 1 } \phi  \right\}, \\
  \end{split}
  							\label{eq.4.30} \\
  p =  p_0  \left( 2 - \kappa  \right)^{ - \kappa / 2 \left( \kappa
    -1 \right) }  \cos^{ - \kappa / \left( \kappa - 1 \right) } \sqrt{
    \kappa - 1 } \phi .
  							\label{eq.4.31}
\end{gather}

  Since the angle \( \phi_* - \phi \) is the angle between the
characteristics and the \( x \) axis, it follows that the line describing
the characteristics is:

\begin{equation}
  y = x \tan \left( \phi_* - \phi \right) + G(\phi).
\label{eq.4.32}
\end{equation}

  The function \( G(\phi) \) is obtained from the following arguments
for a given profile of the curvature \citep{daufm}.  If the equation
describing the shape of the profile is given by the points \( X \text{
and } Y \) where \( Y \! = \! Y(X) \), the velocity of the gas is
tangential to this surface, and so:

\begin{equation}
  \tan \theta = \frac{ \mathrm{d} Y }{ \mathrm{d} X }.
\label{eq.4.33}
\end{equation}

\noindent Now, the equation of the line through the point \( (X,Y) \) which
makes an angle \( \phi_* - \phi \) with the \( x \) axis is:

\begin{equation}
  y - Y = \left( x - X \right) \tan \left( \phi_* - \phi \right).
\label{eq.4.34}
\end{equation}

\noindent Eq.(\ref{eq.4.34}) is the same as eq.(\ref{eq.4.32}) if we set:

\begin{equation}
  G(\phi) = Y - X \tan ( \phi_* - \phi ).
\label{eq.4.35}
\end{equation}

\noindent If we start from a given profile \( Y \! = \! Y(X) \)
then, using eq.(\ref{eq.4.33}) we can find the parametric set of
equations: \( X \! = \! X(\theta) \text{ and } Y \! = \! Y(\theta) \).
Substitution of \( \theta \! = \! \theta(\phi) \) from eq.(\ref{eq.4.27})
or eq.(\ref{eq.4.30}) depending on whether the gas is non--relativistic
or ultrarelativistic, we find \( X \! = \! X(\phi) \text{ and } Y \! = \!
Y(\phi) \).  Substitution of this in eq.(\ref{eq.4.35}) gives the required
function \( G(\phi) \).

  If the shape of the surface around which the gas flows is convex,
the angle \( \theta \) that the velocity vector makes with the \( x \)
axis decreases downstream.  The angle \( \phi - \phi_* \) between the
characteristics leaving the surface and the \( x \) axis also decreases
monotonically.  In other words, characteristics for this kind of flow
do not intersect resulting in a continuous and rarefied flow.

  On the other hand, if the shape of the surface is concave as shown
in Fig.~\ref{fig.2}, the angle \( \theta \) increases monotonically
and so does the angle the characteristics make with the \( x \)
axis.  This means that there must exist a region in the flow in which
characteristics intersect.  The value of the hydrodynamical quantities is
constant for every characteristic line.  This constant however changes
for different non--parallel characteristics.  In other words, at the
point of intersection different hydrodynamical quantities, for example,
the pressure, are multivalued.  This situation cannot occur and results
in the formation of a shock wave.  This shock wave cannot be calculated
from the above considerations, since they were based on the assumption
that the flow had no discontinuities at all because the entropy was
assumed to be constant.  However, the point at which the shock wave
starts, that is point K in Fig.~\ref{fig.2}, can be calculated from
the following considerations.  We can work out the inclination of the
characteristics \( \phi \) as a function of the coordinates \( x \text{
and } y \).  This function \( \phi(x,y) \) becomes multivalued when these
coordinates exceed certain fixed values, say \( x_0 \text{ and } y_0 \).
At a fixed \( x \! = \! x_0 \) the curve giving the value of \( \phi \)
as a function of \( y \) becomes multivalued.  That is, the derivative
\( \left( \partial \phi / \partial y \right)_x \! = \! \infty \), or \(
\left( \partial y / \partial \phi \right)_x \! = \! 0 \).  It is evident
that at the point \( y \! = \! y_0 \) the curve \( \phi(y) \) must lie
in both sides of the vertical tangent, else the function \( \phi(y) \)
would already be multivalued.  This means that the point \( (x_0,y_0)
\) cannot be a maximum, or a minimum of the function \( \phi(y) \) but
it has to be an inflection point.  In other words, the coordinates of
point K in Fig.~\ref{fig.2}  can be calculated from the set of equations
\citep{daufm}:

\begin{equation}
  \left( \frac{ \partial y }{ \partial \phi } \right)_x = 0, \qquad \quad
    \left( \frac{ \partial^2 y }{ \partial \phi^2 } \right)_x = 0.
\label{eq.4.36}
\end{equation}

  When the profile is concave, the streamlines that pass above the point
O in Fig.~\ref{fig.2} pass through a shock wave and the simple wave no
longer exist.  Streamlines that pass below this point seem to be safe
from destruction. However, the perturbing effect from the shock wave KL
influences this region also, and so it is not possible to describe the
flow there as a simple wave.  Nevertheless, since the flow is supersonic,
the perturbing effect of the shock wave is only communicated downstream.
This means that the region to the left of the characteristic PK, which
corresponds to the other set of characteristics emanating from point P,
does not notice the presence of the shock wave. In other words, the
solution mentioned above, in which a simple wave is formed around a
concave profile is only valid to the left of the segment PKL.

\section{Curved jets}
\label{curved-jets}

  Let us now use the results obtained in sections \ref{prandtl-meyer}
and \ref{steady-simple-waves} and apply them to the case of jets that are
curved due to any mechanism, for example the interaction of the jet with
a cloud as was discussed by us in a previous paper \citep{mendoza00},
or due to the ram pressure of the intergalactic gas as in the core of
radio trail sources.

  The greatest danger occurs when the jet forms internal shock waves. This
is because, after a shock, the normal velocity component of the flow to
the surface of the shock  becomes subsonic and the jet flares outward.
Nevertheless, as we have seen in section \ref{steady-simple-waves}, the
shock that forms when gas flows around a curved profile (such as a bent
jet due to external pressure gradients) does not start from the boundary
of the jet.  It actually forms at an intermediate point to the flow.
In other words, it is possible that, if a jet does not bend too much
the intersection of the characteristic lines actually occurs outside
the jet and the flow can curve without the production of internal shocks.

  As we have seen in section \ref{steady-simple-waves} the Mach angle of
the  flow, relativistic and non-relativistic, does not remain constant in
the bend (see for example eq.(\ref{eq.4.27}) and eq.(\ref{eq.4.30})). The
Mach number monotonically decreases as the bend proceeds.

  Eq.(\ref{eq.4.27}) and eq.\eqref{eq.4.30} imply that:

\begin{gather}
  \tan \alpha =  - \mu \cot \mu \left( \alpha + \theta - \phi_* \right).
							\label{eq.4.37} \\
  \intertext{where}
  \mu  \equiv 
    \begin{cases}
      \sqrt{ ( \kappa - 1 ) / ( \kappa + 1 ) }  & \text{ if the gas is
        non--relativistic}, \\
      \sqrt{ \kappa - 1 } & \text{ for an ultrarelativistic gas. }
    \end{cases}
  							\label{eq.4.37a}
\end{gather}

  As was mentioned above, if the jet is sufficiently narrow, it appears
that it can safely avoid the formation of an internal shock.  However,
differentiation of eq.(\ref{eq.4.37}) with respect to the angle the
velocity vector makes with the \( x \) axis, that is the \emph{deflection
angle} \( \theta \), implies that:

\begin{gather}
  \frac{ \mathrm{d} \alpha }{ \mathrm{d} \theta } = \frac{ 1 }{ 2 } \left(
    \Gamma - 1 + \frac{ \Gamma + 1 }{ M^2 - 1 } \right),
							\label{eq.4.38} \\
  \intertext{with}
  \Gamma \equiv
    \begin{cases}
      \kappa  & \text{ if the gas is non--relativistic}, \\
      \kappa / ( 2 - \kappa ) & \text{ for an ultrarelativistic gas.}
    \end{cases}
    							\label{eq.4.38a}
\end{gather}

\noindent The Mach number \( M \) is given by eq.(\ref{eq.2.88}) and
eq.(\ref{eq.2.86}) respectively.  As the Mach number \( M \! \rightarrow
\! 1 \), then the derivative \( \mathrm{d} \alpha / \mathrm{d} \theta
\! \rightarrow \! \infty \). This means that the rate of change of the
Mach angle with respect to the deflection angle grows without limit as
the Mach number decreases and reaches unity.  On a bend, the Mach number
decreases and care is needed, or else the characteristics will intersect
at the end of the curvature.  There is only one special shape for which
this effect is bypassed and this occurs when the increase of \( \theta \)
matches exactly with the increase of \( \alpha \) \citep{courant}, but of
course, this is quite a unique case.  It appears however, that whatever
the thickness of the jet it cannot be bent more than the point at which \(
\mathrm{d} \alpha / \mathrm{d} \theta \) exceeds the rate of change of \(
\theta \) with respect to the bending angle \( \theta \).  In other words,
\( \mathrm{d} \alpha / \mathrm{d} \theta \! \le \! \mathrm{d} \theta /
\mathrm{d} \theta \! = \! 1 \) \citep{icke91}.  From this last inequality
and eq.(\ref{eq.4.38}) a value of the Mach number can be obtained:

\begin{equation}
  M_\star = \frac{ 2 }{ \sqrt{ 3 - \Gamma } }.
\label{eq.4.39}
\end{equation}

\noindent If the Mach number in the jet decreases in such a way that the
value \( M_\star \) is reached, then a \emph{terminal shock} is produced
and the jet structure is likely to be disrupted.  It is important to note
that this terminal shock is weak since \( M \! \gtrsim \! 1 \) and so,
it might not be too disruptive.  Nevertheless, this monotonic decrease
of the Mach number makes the jet flare outwards, even if the terminal
shock is weak.

  Let us now calculate an upper limit for the maximum deflection angle
for which jets do not produce terminal shocks.  In order to do
so, we rewrite eq.(\ref{eq.4.37}) in the following way:

\begin{equation}
  - \theta = \arcsin \frac{ 1 }{ M } + \frac{1}{\mu} \arctan \left\{
    \mu \sqrt{ M^2 - 1 } \right\} - \phi_*
\label{eq.4.40}
\end{equation}

\noindent To eliminate the constant \( \phi_* \) from all our relations,
we can compare the angle \( \theta \) evaluated at the minimum possible
value of the Mach angle \( M \! = \! M_\star \) with \( \theta \)
evaluated at its maximum value \( M \! = \! \infty \).  In other words,
the angle \( \theta_{\text{max}} \) defined as:

\begin{align}
  \theta_{\text{max}} & \equiv \theta ( M \! = \! M_\star ) - \theta
    ( M \! = \! \infty ) 	  		\notag	\\ 
    & = 
    \begin{cases}
      \unit{ 74.21 }{ \degree } & \text{ for the non--relativistic case, } \\ 
      \unit{ 47.94 }{ \degree } & \text{ for an ultrarelativistic gas. }
    \end{cases}
\label{eq.4.41}
\end{align}

\noindent are upper limits to the deflection angle. Jets which bend more
than this limiting value \( \theta_{\text{max}} \) develop a terminal
shock.

\begin{figure}
  \begin{center}
    \psfrag{B}{ \( \alpha \) }
    \psfrag{D}[Bl][Bl][1][90]{ \( \alpha + \mathrm{d} \alpha \) }
    \psfrag{E}{ \( \mathrm{d} \theta \) }
    \psfrag{F}{ \( R \) }
    \psfrag{G}{ \( \mathrm{d} \theta \) }
    \psfrag{H}{ \( D \) }
    \includegraphics[scale=0.53]{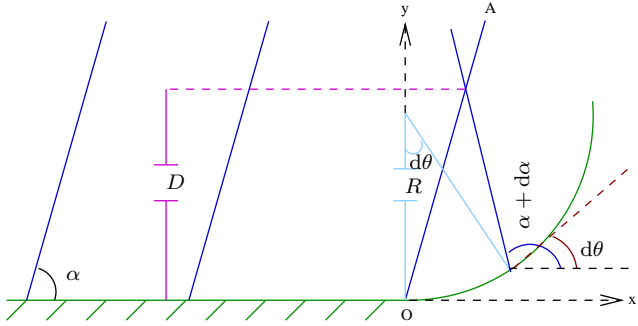}
  \end{center}
  \caption[Sketch of a curved jet that develops a shock at the onset
	   of the curvature]{ Sketch of a curved jet of radius \( D \)
	   that develops a shock at the beginning of the curvature.
	   The curve is assumed to be a circle with radius \( R \),
	   so that this approximation is valid on a sufficiently small
	   region about the onset of the curvature.  The Mach angle of
	   the jet is \( \alpha \) at the left of the characteristic OA
	   that emanates from the point where the bending starts.}
\label{fig.3}
\end{figure}

  This upper limit however, does not mean that the jet is immune
from developing an internal shock if it is bent by a smaller angle.
Indeed, let us suppose that the jet bends and that the curvature it
follows is a segment of a circle as it is shown in Fig.~\ref{fig.3}.
The circle can be considered to be the circle of curvature of the jet's
trajectory formed at the onset of the bending.  According to the figure,
the equation of the characteristic OA that emanates from the point O,
where the curvature starts is:

\begin{equation}
  y = x \tan \alpha
\label{eq.4.42}
\end{equation}

    Once the flow has curved \( \mathrm{d} \theta  \) degrees, the
characteristic at this point is given by:

\[
  \begin{split}
    y & = ( x - R \mathrm{d} \theta ) \tan ( \alpha + \mathrm{d} \alpha +
      \mathrm{d} \phi ) \\
    & \approx x \tan \alpha + x( \mathrm{d} \alpha + \mathrm{d} \theta )
      / \cos^2 \alpha - R \mathrm{d} \theta \tan \alpha,
  \end{split}
\]

\noindent where \(R\) is the radius of curvature of the circular
trajectory.  The intersection of this characteristic  and that given by
eq.(\ref{eq.4.42}) occurs when the \( y \) coordinate has a value:

\[
  D = \frac{ R \sin^2 \alpha }{ 1 + \mathrm{d} \alpha / \mathrm{d} \theta
    }.
\]

\noindent Substitution of eq.(\ref{eq.4.38}) gives \citep{icke91}:

\begin{equation}
  \frac{ D }{ R } = \frac{ 2 }{ \Gamma + 1 } \left( M^2 - 1 \right)
    M^{-4}.
\label{eq.4.43}
\end{equation}

\begin{figure}
  \begin{center}
    \includegraphics[scale=0.5]{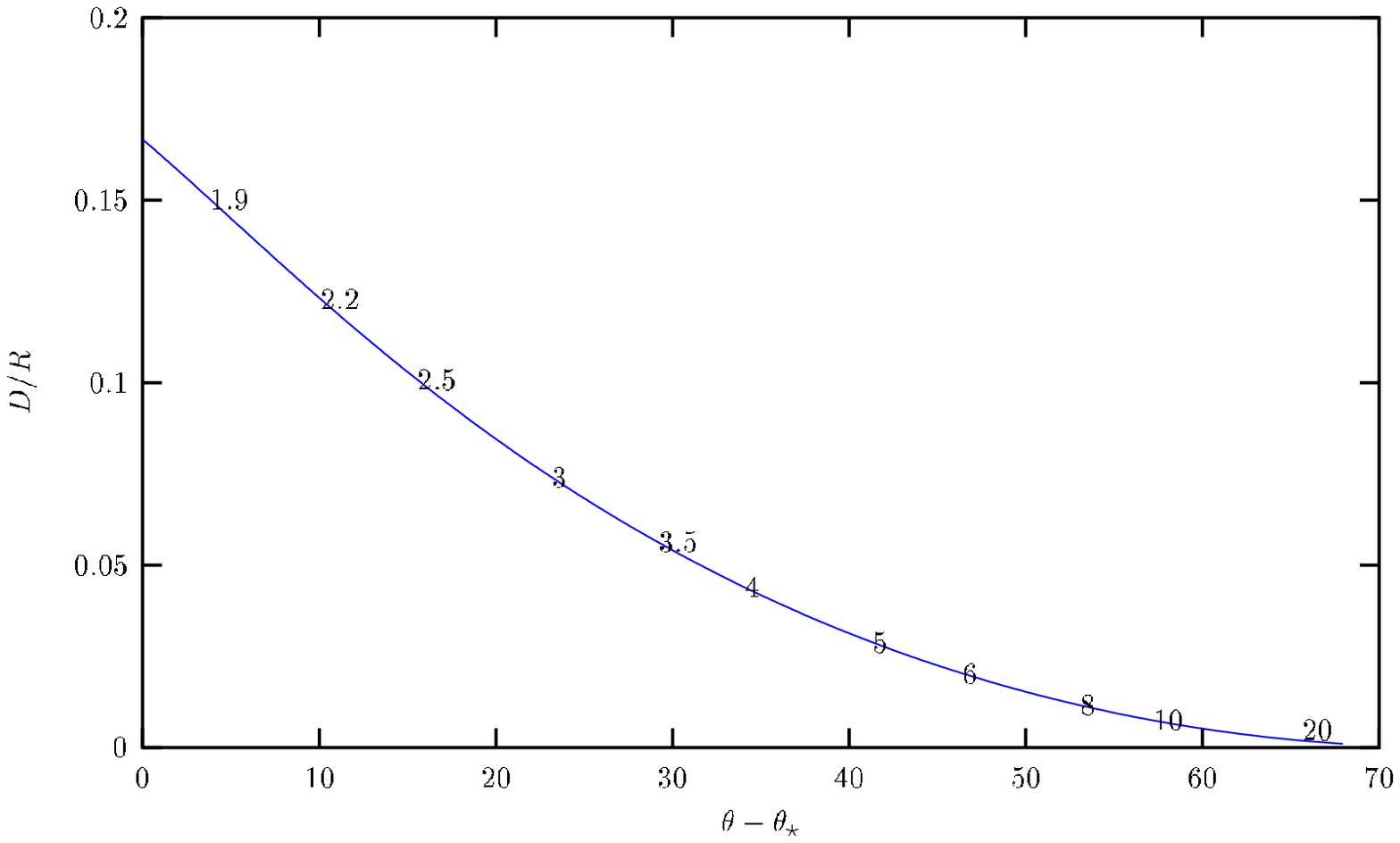}
    \includegraphics[scale=0.5]{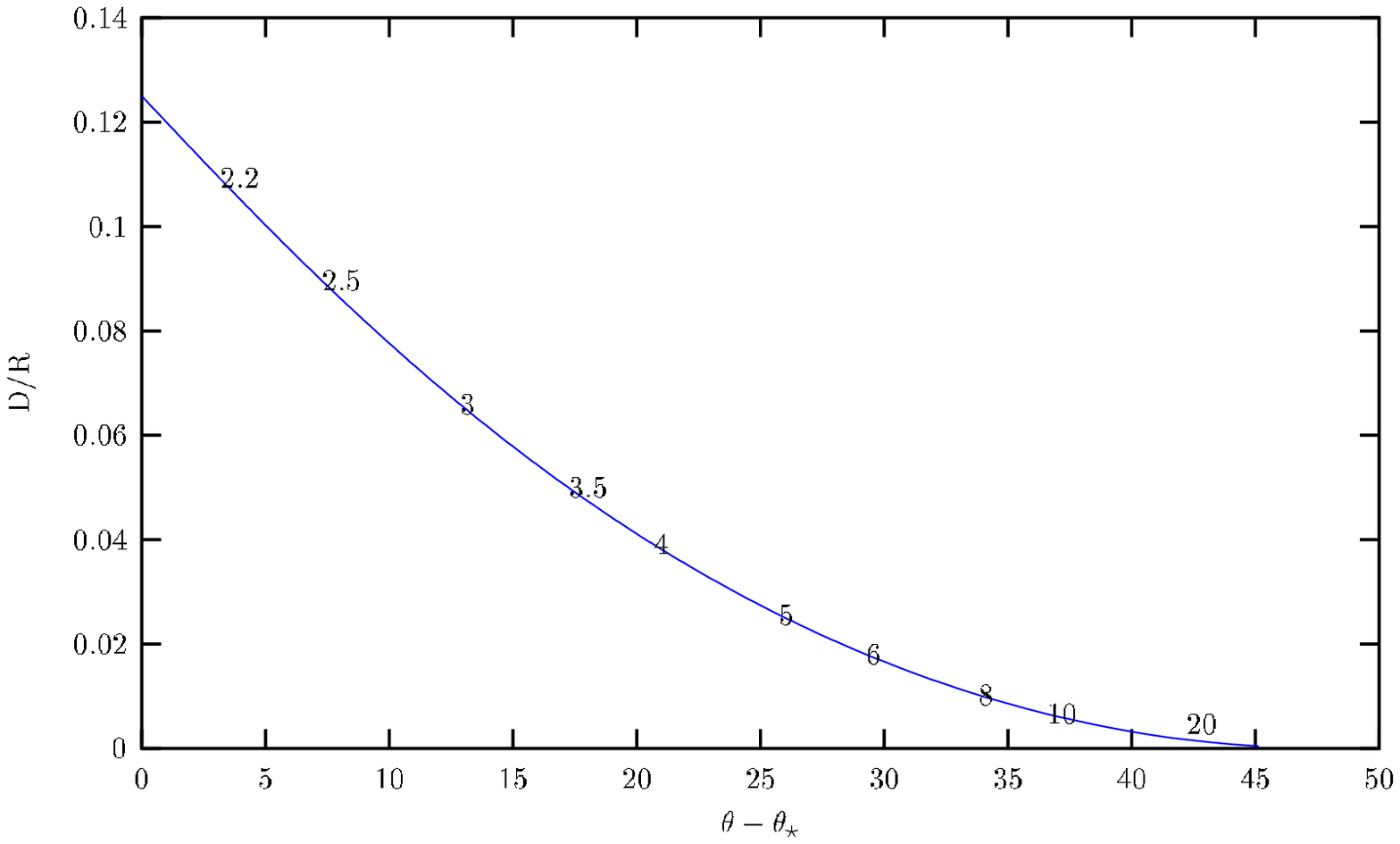}
  \end{center}
  \caption[Generation of a shock at the start of the curvature of a jet]
	  {Plot of the maximum ratio \( D / R \) as a function of
	  the difference \( \theta - \theta_\star \) where \( \theta
	  \) is the deflection angle and \( \theta_\star \) is the
	  maximum bending angle a jet can have in order not to produce
	  a terminal shock.  The plot refers to the points for which a
	  shock at the beginning of the curvature (which was assumed to
	  be a circle) has developed.  Jets with parameters which lie
	  below the curve in any case do not develop any internal shocks
	  at all for this particular circular trajectory.  The plot
	  at the top was calculated using the results in which the
	  gas is non--relativistic and its polytropic index is \( 5 /
	  3 \). The plot at the bottom was made by considering the gas
	  to be ultrarelativistic and relativistic effects in the bulk
	  motion of the flow were taken into account.  For this second
	  plot, the polytropic index was assumed to be \( 4 / 3 \).
	  The numbers in each plot correspond to the values of the Mach
	  number in the flow.}
\label{fig.4}
\end{figure}

  Using eq.(\ref{eq.4.40}) and eq.(\ref{eq.4.43}) it is possible to make a
plot in which two zones separate the cases for jets which develop shocks
at the onset of the curvature, and the ones that do not.  Indeed, we can
plot the ratio of the width of the jet \( D \) to radius of curvature
\( R \) as a function of the difference \( \theta - \theta_\star \)
between the deflection angle \( \theta \) and the maximum deflection
angle \( \theta_\star \! \equiv \! \theta( M_\star ) \), as is shown
in Fig.~\ref{fig.4}.

  Jets for which the ratio \( D / R \) lies below the curve do not
develop any shocks at all.  For example, consider a jet with a given Mach
number for which its ratio \( D / R \) is given.  As the width of the
jet increases (or the radius of curvature of the profile decreases), it
comes a point in which a shock at the onset of the curvature is produced.
In the same way, jets with a fixed ratio \( D / R \) for a given Mach
number which are initially stable, so that they lie below the curve,
can develop a shock at the beginning of the curvature by increasing the
bending angle of the curve.

\section{Discussion}
\label{discussion}

  The relativistic Mach angle is smaller for a given value of the velocity
of the flow than its non--relativistic counterpart as was proved in
section \ref{characteristics}.  This fact is extremely important when
analysing the possibility of the intersection of different characteristics
in a bent jet. This intersection is what gives rise to the creation of
shock waves.  For a relativistic flow, the characteristics, which make
an angle equal to the Mach angle to the streamlines, are always \textsl{
beamed } in the direction of the flow.  Thus, when a jet starts to bend,
the possibilities of intersection between some characteristic line in
the curved jet and the ones before it curves, become more probable than
for their non--relativistic counterpart.

  This difference results in  a severe overestimation of the maximum
bending angle \( \theta_{\text{max}} \) when a non--relativistic
treatment is made to the problem.  For example, \citet{icke91} used the
non--relativistic analysis in the discussion of the generation of internal
shocks due to bending of jets.  Using the non--relativistic equations
described above, but with a polytropic index \( \kappa = 4 / 3 \), the
value for the maximum deflection angle is \( \theta_{\text{max}} \! = \!
\unit{134.16}{\degree} \).  This is much greater than the value of \(
\theta_{\text{max}} \! = \! \unit{74.21}{\degree} \) obtained with a
full relativistic treatment.

  The analysis made by \citet{icke91} is important for jets in which
the microscopic motion of the flow inside the jet is relativistic,
but the bulk motion of the flow is non--relativistic.

  Radio trail sources \citep{begelman84} show considerable bending of
their jets with deflection angles of about \( \unit{90}{\degree} \) in
many cases.  Since the bending is produced by the proper motion of the
host galaxy with respect to the intergalactic medium, the deflection
angle cannot be greater than \( \unit{90}{\degree} \).  The results
presented in eq.\eqref{eq.4.41} show that jets which have a relativistic
equation of state and a bulk relativistic motion of the gas within its
jet, cannot be deflected more than \( \sim \!  \unit{50}{\degree} \).
Since the deflections of radio trail sources are greater than this value,
this result would imply that most radio trail sources should generate
shocks at the end of their curvature.   However, observations \citep[see
for example][and references within]{eilek84,odea85,young91} show that
the velocity of the material of the jets \( \lesssim 0.2\text{--}0.3
\clight \).  Therefore, the bulk motion of the flow is non--relativistic,
despite the fact that the gas inside the jet has a relativistic equation
of state.  As we saw before, this implies that the value of the maximum
angle is \( \theta_{\text{max}} = \unit{134.16}{\degree} \).

  In other words, these type of jets develop a terminal shock if their
jets bend more than \( \sim \! \unit{135}{\degree} \).  This seems to
be the reason why radio trail sources are able to bend so much without
resulting in an internal shock wave that could potentially cause
disruption of its structure.

  In a previous paper \citep{mendoza00}, we discussed the possibility
of a bent jet in the radio galaxy 3C~34 using the observations of
\citet{best97a}.  According to these observations, the radio source lies
more or less in the plane of the sky, and so, if the western radio jet is
curved, this must be of the order of \( \unit{10}{\degree} \).  The value
\( \theta \sim \unit{ 10 }{ \degree } \) is well below the upper limit
of \( \sim \unit{ 50 }{ \degree } \) calculated in eq.\eqref{eq.4.41},
so that no terminal shock would be produced by the deflection
of the jet.  From the lower plot of Fig.~\ref{fig.4} and because the
angle \( \theta_\star \! \ll \! 1 \) for a high relativistic flow, it
follows that if the trajectory of the jet in 3C~34 is circular, then in
order not to produce an internal shock at the onset of the curvature,
the ratio \( D / R \)  has to be less than \( \sim \! 0.08 \).

  In the analysis made above, we have calculated how shocks can be
generated inside a bent jet.  These shocks are special in the sense
that they do not reach the surface boundary of the jet.  Instead they
are generated away from the walls of the jet.  For jets with an
ultrarelativistic equation of state that possess a relativistic bulk
motion, a shock is internally generated if they bend more than \( \sim
\unit{50}{\degree} \).  If their bulk motion is non--relativistic,
the shock is generated when the bending angle is more than \( \sim
\unit{135}{\degree} \).  Jets with a polytropic index of \( 5/3 \) that
move non--relativistically generate a shock if the bending angle exceeds
the value \( \sim \unit{50}{\degree} \).  These angles are only upper
limits and the precise conditions under which a shock is produced have
to be calculated individually.  However using the diagrams presented in
Fig.~\ref{fig.4} and that presented by \citep{icke91} it is possible
to see if a shock is produced at the onset of the curvature.

  All radio sources in which  bendings of radio jets have been observed
appear to satisfy the upper limits discussed above.  So, it seems
that jets are perhaps unstable if an internal shock is generated in
a curvature.  However, various observations and theoretical work in
galactic and extragalactic sources \citep[see for example][and references
within]{canto89,falle93,falle95,komissarov98} show that internal shocks
within a jet are a  good mechanism that, under certain circumstances,
can collimate the jets.

  The question of whether an internal shock will lead to disruption of
the jet is unknown and still a matter of debate. We aim to give to the
problem an answer in a future paper.

\section{Acknowledgements}
  SM would like to thank Paul Alexander for providing ideas to
the problem discussed in this paper while giving a seminar at the
Cavendish Laboratory in Cambridge.  He also thanks support granted by
the Cavendish Laboratory and the Direcci\'on General de Asuntos del
Personal Acad\'emico at the Universidad Nacional Aut\'onoma de M\'exico.

\bibliographystyle{mn2e}
\bibliography{defle}
\label{lastpage}

\end{document}